\documentclass[aps,showpacs,amsmath,amssymb,floatfix,tightenlines,superscriptaddress]{revtex4-2} 

\usepackage[pdftex]{graphicx}       
\usepackage{txfonts}        
\usepackage{mathrsfs}	

\newcommand{\eref}[1]{Eq.~\eqref{#1}}
\newcommand{\fref}[1]{Fig.~\ref{#1}}

\DeclareMathOperator{\var}{var }
\usepackage{hyperref} 
\hypersetup{colorlinks,citecolor=blue,filecolor=blue,linkcolor=blue,urlcolor=blue}

\begin{document}

\title{Critical behaviour of the extended-ballistic transition for pulled self-avoiding walks}

\author{C. J. Bradly} \email{c.bradly@massey.ac.nz}
\affiliation{Dodd-Walls Centre for Photonic and Quantum Technologies, New Zealand Institute of Advanced Study, and Centre for Theoretical Chemistry and Physics, Massey University, Auckland 0632, New Zealand}
\affiliation{School of Mathematics and Statistics, University of Melbourne, Victoria 3010, Australia}
\author{A. L. Owczarek}\email{owczarek@unimelb.edu.au}
\affiliation{School of Mathematics and Statistics, University of Melbourne, Victoria 3010, Australia}

\date{\today}
\begin{abstract}
In order to study long chain polymers many lattice models accommodate a pulling force applied to a particular part of the chain, often a free endpoint.
This is in addition to well-studied features such as energetic interaction between the lattice polymer and a surface. 
However, the critical behaviour of the pulling force alone is less well studied, such as characterizing the nature of the phase transition and particularly the values of the associated exponents.
We investigate a simple model of lattice polymers subject to forced extension, namely self-avoiding walks (SAWs) on the square and simple cubic lattices with one endpoint attached to an impermeable surface and a force applied to the other endpoint acting perpendicular to the surface.
In the thermodynamic limit the system undergoes a transition to a ballistic phase as the force is varied and it is known that this transition occurs whenever the magnitude of the force is positive, i.e.~$f>f_\text{c}=0$.
Using well established scaling arguments we show that the crossover exponent $\phi$ for the finite-size model is identical to the well-known exponent $\nu_d$, which controls the scaling of the size of the polymer in $d$-dimensions.
With extensive Monte Carlo simulations we test this conjecture and show that the value of $\phi$ is indeed consistent with the known values of $\nu_2 = 3/4$ and $\nu_3 = 0.587 597(7)$. Scaling arguments, in turn, imply the specific heat exponent $\alpha$ is $2/3$ in two dimensions and $0.29815(2)$ in three dimensions.
\end{abstract}


\maketitle

\section{Introduction}
\label{sec:Intro}


Lattice models of polymers have been extended to model a huge variety of different physical situations. One well studied addition to lattice polymer models is the application of an external force that acts to deform the polymer chain, possibly with reference to some surface to which it is attached.
These models apply to real applications with atomic force microscopy \cite{Bemis1999,Hansma1996,Haupt1999,Zhang2003} or optical tweezer experiments \cite{Svoboda1994}.
The standard for lattice polymer models is the self-avoiding walk (SAW) and its variants (self-avoiding trails, self avoiding polygons etc.)
Typically, an applied force is included in a SAW model to complement some other interaction or geometric constraint.
A force can be applied to a self-interacting SAW to transition from a collapsed to a stretched phase \cite{Ioffe2010}, or force desorption from an interacting surface \cite{Rensburg2013,Rensburg2016,Rensburg2016b}.
The force can be applied in different ways such as applying it at the midpoint of the polymer \cite{Rensburg2017} or an arbitrary interior point \cite{Bradly2019d}, or it can be applied to move two parallel slabs that confine the polymer \cite{Rensburg2009a}.
A pulling force also appears in models of non-linear lattice polymers such as polygons \cite{Guttmann2018}, stars \cite{Rensburg2019,Bradly2019,Bradly2019c}, or more general lattice animals as well as block copolymers \cite{Rensburg2020}.

The critical properties of this wide range of models have received a lot of attention, but less work has been applied to the behaviour of lattice polymer models that feature {\em only} the applied force \cite{Rensburg2009a,Beaton2015,Ioffe2010}, perhaps because it is among the simplest additions to the bare SAW model.
When a force is applied to the endpoint of a SAW, with the other end terminally attached to some surface, then the SAW behaves ballistically in that its size $R$ from endpoint to endpoint is proportional to the number of steps in the walk, i.e. $R \sim n$.
In the thermodynamic limit of long lengths this occurs for any magnitude of force so there is a critical point at zero force where the size of the walk scales like a standard SAW.
Strictly speaking, it has been shown that the free SAW in the full lattice is sub-ballistic \cite{DuminilCopin2013} but it is generally believed that size scaling of SAW in the half-lattice is the same as the full lattice.
This location of this critical point at $f = 0$ has been proved exactly \cite{Beaton2015,Ioffe2010}, but further properties of this transition are less well understood.
Some previous work \cite{Rensburg2009a} has included Monte Carlo simulation of SAW models with a force applied in slightly different ways but provided limited estimation of critical exponents.

Despite the simplicity of this model and its widespread use as a component in many more complicated models in the literature, a deeper study has not appeared.
In this paper we presume that the transition is not first order and apply standard (finite length) scaling arguments for a continuous transition when a pulling force is applied to the end of a finite-length polymer in two and three dimensions (in fact, in any dimension $d\geq 2$).  
This leads directly to a conjecture for the finite-size crossover exponent $\phi$ and the strength of the transition, indicated by the exponent $\alpha$ which controls the specific heat scaling near the critical point in the thermodynamic limit.
If the size (say radius of gyration) scaling exponent of a free extended polymer in dilute solution is $\nu_d$ in dimension $d$ then our conjecture is that
\begin{align}
\phi =\nu_d, \quad \alpha = 2 - \frac{1}{\nu_d}.
\end{align}
In particular, for two dimensions
\begin{align}
\phi = \nu_2 = \frac{3}{4}, \quad \alpha = \frac{2}{3}
\end{align}
whilst for three dimensions
\begin{align}
\phi = \nu_3 = 0.587 597(7), \quad \alpha = 0.29815(2),
\end{align}
based on the best estimate of $\nu_3$ \cite{Clisby2010}.
It should be noted that $\nu_d \geq 1/2$ for all dimensions $d\geq 2$. 
It is sometimes inferred \cite{Rensburg2009a} that this crossover exponent $\phi$ is close to $1/2$ as it is for the adsorption transition but our numerical evidence discounts this assumption.
Lastly, these numerical values (in particular $\alpha < 1$) indicate that the transition is indeed continuous.
We support this conjecture with strong numerical evidence from Monte Carlo simulations of SAWs on the square and cubic lattices in half spaces where one end is attached to the surface (wall) and a force perpendicular to the wall is applied to the other end.

\section{Model and scaling arguments}
\label{sec:Model}
\begin{figure}[t!]
	\includegraphics[width=0.5\textwidth]{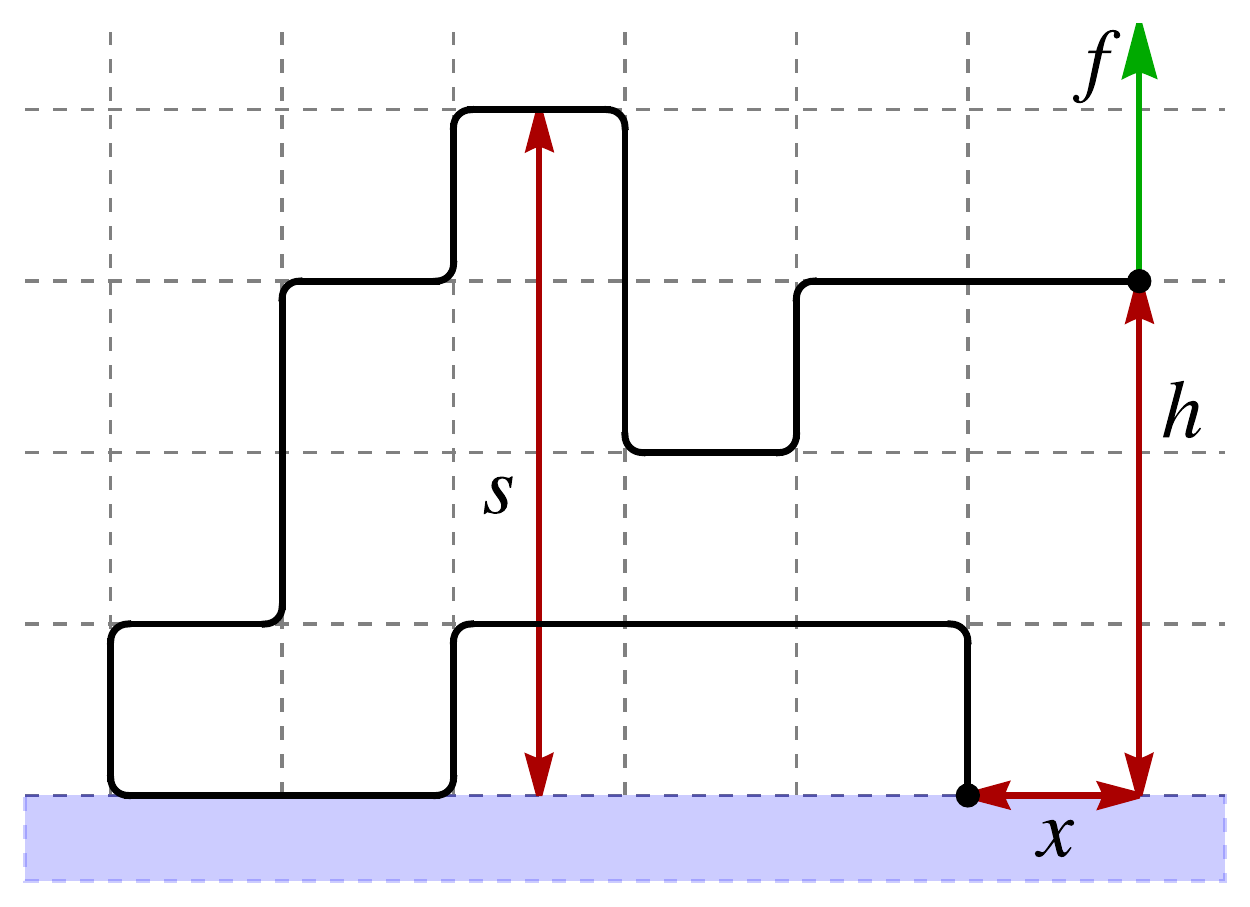}
	\caption{
	A self-avoiding walk on the square lattice with one endpoint attached to an impermeable surface and the other pulled away from that surface by a force $f$.
	The height of the pulled endpoint above the surface is $h$.
	}
	\label{fig:ExamplePulledSAW}
\end{figure}

We consider single polymers in dilute solution modeled as self-avoiding walks (SAWs) on a square or simple cubic lattice.
If the coordinates of each vertex of the walk lie in the space of coordinates defined by  $(x_1,\ldots,x_d)$ with $x_i \in \mathbb{Z}$ for $i=1,\ldots, d$, giving a $d$-dimensional hypercubic lattice $\mathbb{Z}_d$, then we consider walks that are restricted to the half-space defined by $x_d \ge 0$.
One endpoint of the walk is tethered to the surface at the origin and the other endpoint is pulled by a force $f$ perpendicular to the impermeable surface at $x_d = 0$.
This situation is illustrated for two dimensions in \fref{fig:ExamplePulledSAW}.
Let the coordinates of the vertices be denoted as $(x_{1,j},\ldots,x_{d,j})$ for $j=0,\ldots,n$.  We denote by $h=x_{d, n}$ the height of the pulled endpoint and by $a_{nh}$ the number of such walks on the lattice that are of length  (number of steps) $n$.
The number of {\em all} walks of length $n$ confined to one half-space is $\sum_h a_{nh}$ and it is known that this quantity has the same large $n$ limit as unrestricted SAWs on a hypercubic lattice \cite{Hammersley1982}.
That is, $\lim_{n \to \infty} \tfrac{1}{n} \log \sum_h a_{nh} = \log \mu_d$, where $\mu_d$ is the growth constant of the $d$-dimensional hypercubic lattice \cite{Whittington1975}.

If the height of the endpoint is due to a force $f$ then we associate a Boltzmann factor $y^h$ where $y = \exp(f/k_\text{B} T)$.
The canonical partition function of such walks of length $n$ with endpoint at height $h$ is
\begin{equation}
	Z_n(f) = \sum_{h} a_{nh} \, y^h.
	\label{eq:Partition}
\end{equation}
The (reduced) finite-length free energy is
\begin{equation}
	\lambda_n(f) = - \frac{1}{n} \log Z_n(f).
	\label{eq:FreeEnergyFinite}
\end{equation}
and we consider the average height per step
\begin{equation}
	h_n(f) = \frac{\langle h \rangle (f)}{n} = \frac{1}{n} \frac{\sum_{h} h \,a_{nh} \, y^h}{Z_n(f) } = - \frac{d \lambda_n(f)}{d f}
	\label{eq:AverageHeight}
\end{equation}
and its variance
\begin{equation}
	c_n(f) = \frac{\var(h)}{n} = \frac{\langle h^2 \rangle - \langle h \rangle^2}{n} =  \frac{d h_n(f)}{d f},
\label{eq:SpecificHeat}
\end{equation}
which we also call the specific heat since it can be expressed as the second derivative of the free energy.

\subsection{Thermodynamic scaling}

In the thermodynamic, or long-length, limit the free energy is $\lambda(f) = \lim_{n \to \infty} \lambda_n(f)$ and it is known that $\lambda(f)$ is singular at exactly $f_c = 0$, or equivalently $y_c = 1$, which is the location of the extended-ballistic transition \cite{Beaton2015,Ioffe2008,Ioffe2010}.
This holds for all dimensions $d \ge 2$.
Considering the thermodynamic limits
\begin{equation}
	H(f) = \lim_{n\rightarrow\infty} h_n(f)
	\label{eq:AverageHeightLimit}
\end{equation}
and 
\begin{equation}
	C(f) =  \lim_{n\rightarrow\infty} c_n(f)
	\label{eq:SpecificHeatLimit}
\end{equation}
then the transition is signified by the behaviour
\begin{equation}
	H(f) = C(f) = 0 \mbox{ for } f \leq 0
	\label{eq:PushedPhaseOrderParameter}
\end{equation}
and
\begin{equation}
	H(f) > 0  \mbox{ and } C(f) > 0 \mbox{ for } f > 0
	\label{eq:PulledPhaseOrderParameter}
\end{equation}
the average height per step can be considered an order parameter for the transition.

The definition of the critical exponent $\alpha$ is
\begin{equation}
C(f) \sim \left( f - f_\text{c} \right)^{-\alpha}  \mbox{ as } f \rightarrow f_\text{c},
	\label{eq:alpha}
\end{equation}
and a continuous transition is signified by $\alpha < 1$.
The standard crossover scaling Ans\"atze for a continuous transition near a critical point $f_\text{c}$ at large length $n$ are
\begin{equation}
	c_n(f) \sim n^{\alpha \phi} \tilde{C}\left( \left[ f - f_\text{c} \right] n^\phi \right) \mbox{ as } n\to \infty \mbox{ as } f \to f_\text{c}
	\label{eq:SpecificHeatCrossover}
\end{equation}
where $\phi < 1$ is the crossover exponent, and
\begin{equation}
	h_n(f) \sim n^{\alpha \phi - \phi} \tilde{H}\left( \left[ f - f_\text{c} \right] n^\phi \right) \mbox{ as } n \rightarrow \infty \mbox{ as } f \rightarrow f_\text{c}.
	\label{eq:HeightCrossover}
\end{equation}
These two expressions are simply related by differentiation. The specific heat is consistent with the definition of $\alpha$ through (\ref{eq:alpha}) by requiring that the scaling function $\tilde{C}(z) \sim z^\alpha$ for $z\rightarrow\infty$.

By standard tricritical scaling arguments \cite{Brak1993} it follows from the crossover scaling expressions Eqs.~\ref{eq:SpecificHeatCrossover} and \ref{eq:HeightCrossover} and the free energy scaling of $\frac{1}{n}$ that there is a relation between $\phi$ and $\alpha$
\begin{equation}
	2 - \alpha = \frac{1}{\phi},
	\label{eq:HyperscalingRelation}
\end{equation}
so that when $f = f_\text{c} = 0$ we have
\begin{equation}
	h_n(0) \sim H_0 \, n^{\phi - 1} 
	\label{eq:HeightScalingAtF0}
\end{equation}
and 
\begin{equation}
	c_n(0) \sim C_0 \, n^{2\phi - 1}.
	\label{eq:SpecificHeatScalingAtF0}
\end{equation}
The constants $H_0$ and $C_0$ are the values of the scaling functions $\tilde{H}$ and $\tilde{C}$ at $f = 0$.

Now at $f = 0$ given that the height of the polymer end will scale in proportion to the size of the polymer generally given the polymer has a free end it would also be expected that
\begin{equation}
	\langle h \rangle (0) \sim A \; n^{\nu_d},
	\label{eq:NoForceSizeScaling}
\end{equation} 
for some constant $A$ and where $\nu_d$ is the isotropic free $d$-dimensional size scaling exponent: in two dimensions it is predicted that $\nu_2=3/4$ and in three dimensions the best estimate is $\nu_3=0.587 597(7)$ \cite{Clisby2010}.
Hence, by comparing \eref{eq:NoForceSizeScaling} with \eref{eq:HeightScalingAtF0} (recall that $h_n = \langle h \rangle / n$) we can deduce a particular scaling relation for this pulling-pushing transition
\begin{equation}
	\phi = \nu_d.
	\label{eq:CriticalScalingRelation}
\end{equation}
Although this relation is simple, it will hold in all models with a force that changes from pulling to pushing.


Away from the critical point where the endpoint is subject to a pulling or pushing force, the value of $\nu$ can change and we should look at some other size-related quantities.
Let the end point of our walks be at $(x_{1,n},h)$ or $(x_{1,n},x_{2,n},h)$ for two and three dimensions respectively, and let the walk fit into a slab or slit of no less than $s= \max\{x_{d,j}, j = 0,\ldots, n\}$ units high (that is, all sites of the walk have height less than and or equal to $s$. 
We note $s \geq h$, and 
\begin{equation}
 r_x = \left [ \sum_{i=1}^{d-1} x_{i,n}^2 \right ]^{1/2}
\label{eq:R2Parallel}
\end{equation}
is shorthand for the distance from the origin to the projection of the endpoint onto the impermeable surface, generalised for all dimensions $d$.
Similar to the discussion above, we define size exponents $\nu^{(h)}$, $\nu^{(x)}$ and $\nu^{(s)}$ at fixed $f$
\begin{subequations}
\begin{align}
	\langle h \rangle (f) &\sim H_0\,  n^{\nu^{(h)}}\,,\\
	\langle r_x \rangle (f) &\sim X_0 \, n^{\nu^{(x)}}\,,\\
	\langle s \rangle (f) &\sim S_0 \, n^{\nu^{(s)}}\,.
\end{align}
\label{eq:SizeQuantities}
\end{subequations}
The expected values of these exponents in each phase and at the critical point is summarised in Table \ref{tab:SizeExponents}.
In the extended phase, where $f < 0$, or equivalently $y < 1$, the endpoint is pushed into the surface, that is, $\langle h \rangle = 0$ and hence $\nu^{(h)} = 0$.
However, the pushing force is local to the endpoint and the size of the rest of the polymer is also characterised by the same exponents $\nu_d$.
As $f \to -\infty$, or equivalently $y \to 0$, the pushed endpoint is effectively fixed in the surface, equivalent to a self-avoiding {\em loop} \cite{Rensburg2015}.
The ballistic phase, $f > 0$, is so named because it is characterised by the behaviour $h_n \sim O(1)$ and thus it is known that $\nu^{(h)} = 1$ for SAWs \cite{Ioffe2008}.
Since $s$ is an alternate measure of the extension of the polymer when pulled away from the surface, we expect $\nu^{(s)} = 1$ as well.

For $\nu^{(x)}$ in the ballistic phase note that the polymer is stretched so the component of the endpoint position parallel to the surface is small.
Intuitively, the endpoint is the unique highest point so it is like a simple random walk in a $d-1$ dimensional plane parallel to the impermeable surface and thus we have $\nu^{(x)} = 1/2$.
Another approach is to assume that a stretched polymer in the ballistic phase corresponds to a certain directed self-avoiding walk.
In particular, a free SAW without pulling but that only allows steps parallel or away from the surface. 
In this case the size scaling orthogonal to the directed axis is $n^{1/2}$ \cite{Redner1983}.

\begin{table}[t!]
	\begin{tabular}{l|c|c|c|c}
	\hline
	Phase &  $f$ & $\nu^{(h)}$ & $\nu^{(x)}$ & $\nu^{(s)}$ \\
	\hline\hline
	Extended (pushed) & $f<f_c$ &  0 & $\nu_d$ &  $\nu_d$ \\
	\hline
	Critical point & $f=f_c$ &  $\nu_d$ & $\nu_d$ &  $\nu_d$ \\
	\hline
	Ballistic (pulled) & $f>f_c$ & 1 &$\frac{1}{2}$ & 1 \\
	\end{tabular}
\caption{Size exponent values for the phases in our model.
The exponents $\nu^{(h)}$, $\nu^{(x)}$ and $\nu^{(s)}$ are defined by \eref{eq:SizeQuantities} and characterise the scaling of the endpoint height $h$, endpoint horizontal extensions $x$ and total height $s$, respectively. The critical point is at $f_\text{c} = 0$, or equivalently $y_\text{c} = 1$.}
\label{tab:SizeExponents}
\end{table}

\subsection{Simulation details}
Walks are simulated using the flatPERM algorithm \cite{Prellberg2004}, an extension of the pruned and enriched Rosenbluth method (PERM) \cite{Grassberger1997}. 
The simulation works by growing a walk on a given lattice up to some maximum length $N_\text{max}$. 
At each step the number of bulk interactions $m$ and straight segments $s$ are calculated and the cumulative Rosenbluth \& Rosenbluth weight \cite{Rosenbluth1955} is compared with the current estimate of the weights of all samples $W_{nh}$. 
If the current state has relatively low weight the walk is `pruned' back to an earlier state. 
On the other hand, if the current state has relatively high weight, then microcanonical quantities $h$ are measured and $W_{nh}$ is updated. 
The state is then `enriched' by branching the simulation into several possible further paths (which are explored when the current path is eventually pruned back). 
When all branches are pruned a new iteration is started from the origin.
FlatPERM enhances this method by altering the prune or enrich choice such that the sample histogram is flat in the microcanonical parameters $n$ and $h$. 
Further improvements are made to account for the correlation between branches that are grown from the same enrichment point, which provides an estimate of the number of effectively independent samples. 
The main output of the simulation are the weights $W_{nh}$, which are an approximation to the athermal density of states $a_{n,h}$ in \eref{eq:Partition}, for all $n\le N_\text{max}$. 

Once the simulation is finished thermodynamic quantities are determined by specifying the Boltzmann weight $y$ and using the weighted sum
\begin{equation}
    \langle Q \rangle_n = \frac{\sum_{h} Q_{nh} y^h W_{nh}}{\sum_{h} y^h W_{nh}}.
    \label{eq:FPQuantity}
\end{equation}
In particular, we calculate $h_n$ and its variance $c_n$ according to Eqs.~\ref{eq:AverageHeight} and \ref{eq:SpecificHeat} as well as $\langle x \rangle$ and $\langle s \rangle$.
Similarly we also calculate the third derivative of the free energy $\lambda_n(f)$ 
\begin{equation}
	t_n(f) = \frac{d^3\lambda_n(f)}{df} = \frac{\langle h^3 \rangle - 3\langle h \rangle \langle h^2 \rangle + 2\langle h \rangle^3}{n}.
\label{eq:ThirdDerivativeSimulation}
\end{equation}

We use the flatPERM algorithm to simulate walks up to length $N_\text{max} = 1024$.
Because this model is simple we can afford to run very long simulations to obtain good statistics.
For both 2D and 3D models we have $1.2 \times 10^{7}$ iterations, obtaining $5.2 \times 10^{12}$ samples for 2D walks and $6.3 \times 10^{12}$ samples for 3D walks.
For each case or simulation mentioned in this work the results are comprised of a composite of ten independent simulations in order to obtain some measure of statistical error.
In the following section this error falls within the line or marker width of the plots, or is otherwise marked.
We ran separate smaller simulations ($N_\text{max} = 512$) to obtain data for $\langle s \rangle$ in order to verify scaling predictions.

\begin{figure*}[t!]
	\includegraphics[width=\textwidth]{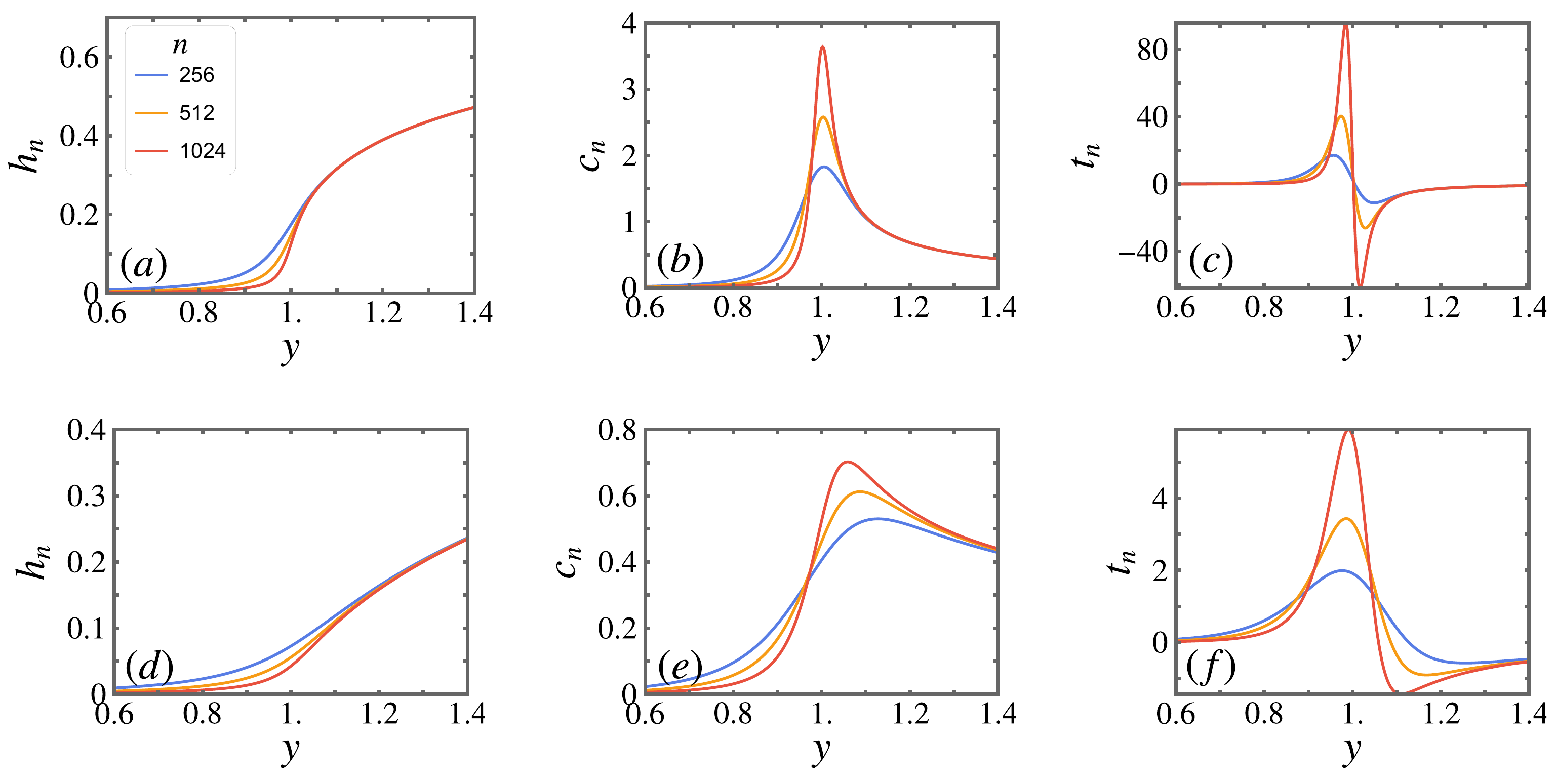}
	\caption{
	Thermodynamic quantities for pulled SAWs in 2D (a-c) and 3D (d-f), for several different lengths $n$. From left to right are the average extension above the surface $h_n$, the specific heat $c_n$, and the third derivative of the free energy $t_n$.
	}
	\label{fig:Thermo}
\end{figure*}

\section{Results}
\label{sec:Results}

We first show in \fref{fig:Thermo} several thermodynamic quantities for the 2D walks (a-c) and 3D walks (d-f) as functions of the Boltzmann weight $y = e^{f/k_B T}$.
The left plots show the average extension above the impermeable surface, $h_n$, the specific heat $c_n$ and the third derivative of the free energy $t_n$.
All plots show example curves for $n = 256,512,1024$.
There is a clear signal of a transition near $y=1$ ($f=0$), most notably as a peak in $c_n$ that grows with $n$.
The behaviour of $h_n$ is suggestive of a continuous transition.

\begin{table}[b!]
\begin{tabular}{l|c|c|c|c}
lattice & $\quad$ & $h_n$ & $c_n$ & $t_n$ \\ \hline\hline
squ & $\alpha$ 	& $0.6630 \pm 0.0024$ & $0.6673 \pm 0.0031$ & $0.638 \pm 0.020$ 		\\
		& $\phi$ 		& $0.7479 \pm 0.0009$ & $0.7504 \pm 0.0011$ & $0.734 \pm 0.007$ 		\\ \hline
sc	& $\alpha$	&	$0.2997 \pm 0.0029$ & $0.3038 \pm 0.0033$ 	& $0.300 \pm 0.009$		\\
		& $\phi$ 		& $0.5881 \pm 0.0009$ & $0.5896 \pm 0.0010$ & $0.5882 \pm 0.0026$ 	
\end{tabular}
\caption{Critical exponents $\alpha$ and $\phi$ estimated from the scaling of thermodynamic quantities at the critical point $f_\text{c}$ for pulled SAWs on the square and simple cubic lattices.}
\label{tab:Exponents}
\end{table}

In order to verify the type of transition we estimate the critical exponents using a correction-to-scaling method.
That is, for a quantity $\langle Q \rangle_n$ that has power-law leading order behaviour $n^b$ for some expected exponent $b$, we fit the data to
\begin{equation}
	\langle Q \rangle_n \sim A \, n^b \left ( 1 + C \, n^{-\Delta} \right ),
\label{eq:CorrectionToScaling}
\end{equation}
where $A$ and $C$ are fitting parameters and $\Delta$ is a correction to scaling parameter.
There are a range of corrections to scaling methods that have been seen in the literature where $n^{-\Delta}$ is only the first correction term of an infinite series.
We do not have the necessary numerical precision to estimate $\Delta$ independently for this model and the results only weakly depend on its value as long as $\Delta = O(1)$.
Hence, for our numerical simulations we use a single term with $\Delta = 1/2$ for three dimensions and $\Delta = 1$ for two dimensions, which are typical values seen in the literature \cite{Clisby2010,Caracciolo2005,Guttmann2001,Nienhuis1982,Shannon1996}.
All reported results for exponents $\phi$ and $\alpha$ are derived from a model of the form of \eref{eq:CorrectionToScaling}, although in most cases the effect of adding the correction-to-scaling term is weak, usually in the third decimal place.

Three estimates of the exponents have been made using correction-to-scaling models with the following leading order exponents.
First, we fit the average extension at the critical point $h_n(f_\text{c})$ with leading order exponent $\phi-1$, as per \eref{eq:HeightScalingAtF0}.
Second, we fit the specific heat at the critical point $c_n(f_\text{c})$ with leading order exponent $2\phi-1$, as per \eref{eq:SpecificHeatScalingAtF0}.
Third, we fit the critical point values of $t_n(f_\text{c})$.
By similar scaling arguments to the previous section the peak value of $t_n$ is expected to scale with leading-order exponent $(\alpha+1)\phi = 3\phi - 1$.
This can be obtained by differentiating \eref{eq:SpecificHeatCrossover} and using the standard relation \eref{eq:HyperscalingRelation}.
The values of $\phi$ determined from these fits are reported in Table \ref{tab:Exponents}.
In all cases the value of $\alpha$ is then derived from the relation in \eref{eq:HyperscalingRelation} and is not an independent estimate, but is reported for ease of comparison.

These estimates are consistent with $\alpha \approx 2/3$ and $\phi = 3/4$ for the square lattice, confirming that $\phi = \nu_2$ in this case.
For the simple cubic lattice the values of $\phi$ are close to the expected $\phi = \nu_3 \approx 0.587597(7)$, or equivalently by the hyperscaling relation \eref{eq:HyperscalingRelation} $\alpha = 0.29815(5)$.
For both two and three dimensions these exponents confirm that the extended-ballistic transition is continuous.

\begin{figure}[t!]
	\includegraphics[width=0.8\textwidth]{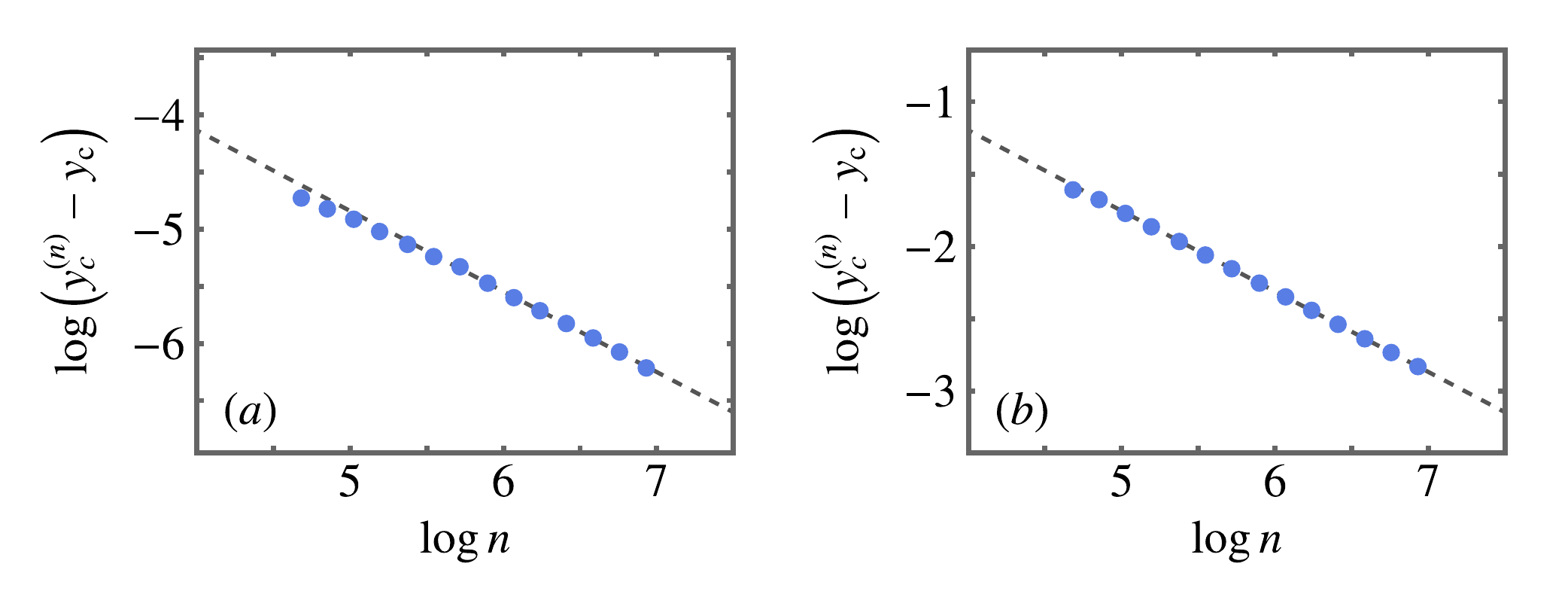}
	\caption{
	The peak values of $c_n$ as an estimate for the location of the transition $y_c^{(n)}$ for (a) the square lattice and (b) the simple cubic lattice.
	}
	\label{fig:TcFits}
\end{figure}
\begin{figure}[t]
	\includegraphics[width=0.8\textwidth]{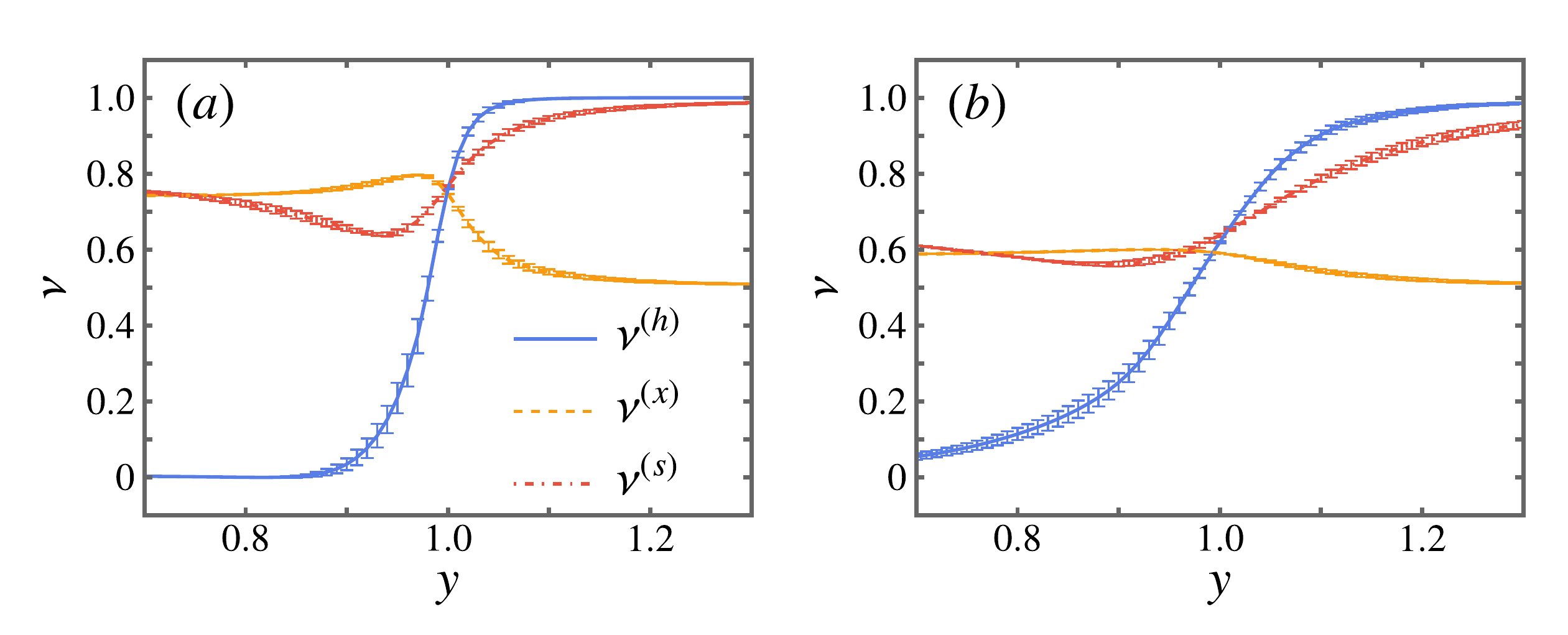}
	\caption{
	The size exponent $\nu$ from several quantities across the ballistic transition for (a) 2D and (b) 3D. 
	At the critical point $y_\text{c} = 1$ the exponents coincide with value $\nu_d$.
	Estimates for $\nu^{(h)}$ and $\nu^{(x)}$ are from simulations up to $n_\text{max}=1024$ while estimates for $\nu^{(s)}$ are from additional simulations up to $n_\text{max}=512$.
	}
	\label{fig:R2}
\end{figure}

The location of the extended-ballistic transition is rigorously known \cite{Beaton2015,Ioffe2010} so we have the luxury of immediately looking at the scaling of thermodynamic quantities at exactly $y_\text{c} = 1$ ($f_\text{c} = 0$).
This saves a lot of additional work to accurately locate the critical point, for which one needs some easily identifiable signal of the transition.
In general for a finite-size system, the critical point deviates from the long chain limit as
\begin{equation}
	y_c^{(n)} - y_\text{c} \sim n^{-\phi},
\label{eq:TcDrift}
\end{equation}
where the crossover exponent $\phi$ appears again.
An obvious choice for $y_c^{(n)}$ is the {\em location} of the peak value of the specific heat $c_n^{(\text{peak})}$, which occurs near, but slightly above, the transition point.
In the thermodynamic limit the specific heat at the critical point diverges and so $c_n(0)$ should also diverge with $n$ if the crossover exponent $\phi > 1/2$.
This is the case for our model and one could even approximate $c_n(0) \approx c_n^{(\text{peak})}$.
By comparison, locating the critical point is a significant issue for other systems, e.g.~polymer adsorption where $\phi \leq 1/2$ and so the peaks of the specific heat are not a good indicator of the transition \cite{Rensburg2004,Bradly2018}.
Thus, while our earlier estimates of the exponents are justified, it is nevertheless worth comparing the known exact value of the transition point to estimates from the simulations using the peak values of $c_n$.
This also provides a further estimate of $\phi$.

In \fref{fig:TcFits} we show a log-log plot of $y_c^{(n)} - y_\text{c} = c_n^{(\text{peak})} - 1$.
First, it is clear that in both cases the long-length limit $y_\text{c} = 1$ is good.
The dashed lines are least-squares fits to a simple power law (without correction to scaling) to the larger values of $n$.
Particularly for the the square lattice, this is the case where we see the largest finite-size effect at small $n$.
Without the correction to scaling we find $\phi = 0.705 \pm 0.008$ for the square lattice and $0.557 \pm 0.001$ for the simple cubic lattice.
When a correction-to-scaling model is used with expected leading order exponent $-\phi$ according to \eref{eq:TcDrift}, we find improved estimates of $\phi = 0.76 \pm 0.05$ for the square lattice and $\phi = 0.567 \pm 0.003$ for the simple cubic lattice.
These values closely match the estimates from the scaling of thermodynamic quantities presented above and thus are again in good agreement with the relation $\phi = \nu_d$.
Our results disagree with $\phi \approx 1/2$ found in \cite{Rensburg2009a}, which also estimated $\phi$ from the shift of the critical point at finite lengths.


Lastly, we look at the size scaling of the walks via the quantities in \eref{eq:SizeQuantities}.
Additional simulations with smaller maximum length $n_{\max} = 512$ were run to obtain data for $\langle s \rangle$.
Estimates for the scaling exponents of these quantities were obtained by simple power-law fits (without finite-size corrections to scaling) for a range of values of $y$ spanning the extended-ballistic transition.
In \fref{fig:R2} we plot the exponents $\nu^{(h)}$, $\nu^{(x)}$ and $\nu^{(s)}$, for the square (a) and simple cubic (b) lattices.
Broadly, the values of these exponents match what is outlined in Table \ref{tab:SizeExponents}.
In particular, the values of the different exponents coincide at the critical point $y_\text{c} = 1$ where it is expected that all exponents have value $\nu_d$.
However, the estimates for the simple-cubic lattice show a larger spread compared to the square.
We expect that this is due to finite size effects, particularly in the estimate of $\nu^{(s)}$ that was obtained from simulations with a smaller $n_\text{max}$.

\section{Conclusion}
\label{sec:Conclusion}

We have investigated the critical behaviour of a pulled SAW near the known extended-ballistic transition at zero force $f_\text{c} = 0$ ($y_\text{c} = 1$).
With standard scaling analysis we find that the exponent $\phi$ governing the crossover from the finite size behaviour to the thermodynamic limit is the same as the size exponent $\nu$.
The exponent $\alpha$ governing the the scaling of the specific heat follows from standard scaling relations. We have verified these scaling relations through independent estimation of exponents in different quantities.
In two and three dimensions $\alpha<1$, indicating that the extended-ballistic transition is a continuous transition.
Although this model is simple, the scaling relation and other critical properties of the transition have not been explicitly stated in the literature, despite the widespread use of a pulling force in lattice polymer models.
Monte Carlo simulation showed that this result is verified in the scaling of thermodynamic quantities at the critical point.
It is also verified in the drift of the finite-size critical point $y_c^{(n)}$ from the thermodynamic limit $y_\text{c} = 1$, as represented by a typical signature, namely the location of the specific heat peak.
Furthermore, we looked at the size exponent in both the extended and ballistic phases by way of several quantities that are related to the extension of the polymer away from the surface.
The scaling of these quantities shows consistent estimations of the size exponent $\nu$.

\begin{acknowledgments}
The authors thank Stu Whittington for suggesting this problem to us and for helpful comments.
This research was supported by The University of Melbourne’s Research Computing Services and the Petascale Campus Initiative.
\end{acknowledgments}

\section*{Research data}
The raw simulation data used in this article is available on request to the authors. 

\bibliography{polymers_master}{}

\end{document}